\documentstyle[epsfig,12pt]{article}
\begin{document}


\def\etal{{\it et.~al.}}
\def\ie{{\it i.e.}}
\def\eg{{\it e.g.}}

\def\be{\begin{equation}}
\def\ee{\end{equation}}
\def\bea{\begin{eqnarray}}
\def\eea{\end{eqnarray}}
\def\bean{\begin{eqnarray*}}
\def\eean{\end{eqnarray*}}
\def\bary{\begin{array}}
\def\eary{\end{array}}
\def\bi{\bibitem}
\def\bit{\begin{itemize}}
\def\eit{\end{itemize}}

\def\lan{\langle}
\def\ran{\rangle}
\def\lra{\leftrightarrow}
\def\la{\leftarrow}
\def\ra{\rightarrow}
\def\dash{\mbox{-}}

\def\re{\rm Re}
\def\im{\rm Im}
\def\eps{\epsilon}
\def\sg{\tilde g}
\def\sb{\tilde b}
\def\st{\tilde t}
\def\vk{\bf k}
\def\vp{\bf p}
\def\ua{\uparrow}
\def\da{\downarrow}


\large \centerline {\bf Heavy bottom squark mass in}
\centerline {\bf the light gluino and light bottom squark
scenario \footnote{Enrico Fermi Institute preprint EFI 03-27,
hep-ph/0306022, to be submitted to Phys. Lett. B.}}
\normalsize

\bigskip

\centerline {Zumin Luo~\footnote{zuminluo@midway.uchicago.edu} and
Jonathan L. Rosner~\footnote{rosner@hep.uchicago.edu}} \centerline
{\it Enrico Fermi Institute and Department of Physics}
\centerline{\it University of Chicago, 5640 S. Ellis Avenue,
Chicago, IL 60637}

\begin{quote}
Restrictive upper bounds on the heavy bottom squark mass when the gluino and
one bottom squark are both light are based on the predicted reduction of
$R_b$ (the fraction of $Z$ hadronic decays to $b \bar b$ pairs) in such a
scenario.  These bounds are found to be relaxed by the process $Z \to
b\bar{\sb}\sg/{\bar b}\sb\sg$, which may partially compensate for the
reduction of $R_b$.  The relaxation of bounds on the top squark and the
scale-dependence of the strong coupling constant are also discussed.
\end{quote}

\bigskip
\noindent
PACS Categories: 11.30.Pb, 12.60.Jv, 13.38.Dg, 14.80.Ly

\bigskip

As a promising candidate for physics beyond the Standard Model (SM), the
minimal supersymmetric standard model (MSSM) has been intensively studied
in the past years. Recently there has been some interest \cite{Berger:2000mp,
refs, Cao:2001rz, Cho:2002mt, Baek:2002xf, Cheung:2002na, Chiang:2002wi,
Malhotra:2003da, Luo:2003zj} in a MSSM scenario with a
light gluino and a light bottom squark. It was first proposed by Berger
{\it et al.} \cite{Berger:2000mp} to explain the excess of the cross
section for bottom quark production at hadron colliders. In this scenario,
the gluino has mass $m_{\sg}=$ 12 -- 16 GeV and the light bottom squark
has mass $m_{\sb}=$ 2 -- 5.5 GeV. The other bottom squark $\sb'$, an
orthogonal mixture of $\sb_L$ and $\sb_R$ to the light bottom squark, is
assumed to be heavy, and so are all other supersymmetric 
(SUSY) particles. We follow the convention in Ref.~\cite{Berger:2000mp} to
define
\be \left( \begin{array}{c} \sb \\ \sb' \end{array} \right) =
\left( \begin{array}{c c} \cos \theta_{\sb} & \sin \theta_{\sb} \cr
                        - \sin \theta_{\sb} & \cos \theta_{\sb}
\end{array} \right)
\left( \begin{array}{c} \tilde b_R \\ \tilde b_L \end{array}
\right)~~~. \ee
In this scenario the $Z$ decay width $\Gamma_Z$ is reduced by SUSY-QCD
corrections to the $Z b {\bar b}$ vertex \cite{Cao:2001rz}. However, some
new decay processes contribute positively to $\Gamma_Z$ and may compensate
for the reduction. Apart from $Z \to \sg \sg$, which was studied in 
Ref.~\cite{Luo:2003zj} and found to have a decay width of order 0.1 MeV,
there exist two more important decays, $Z \to b\bar{\sb}\sg/{\bar 
b}\sb\sg$ and $Z \to q{\bar q}\sg\sg$. The first process is $\sim \alpha
\alpha_s$ at the tree level and has a decay width of 1.9 -- 5.9 MeV
depending on the sign of $\sin 2\theta_{\sb}$ and the mass of the gluino
\cite{Cheung:2002na}. The second process is $\sim \alpha \alpha_s^2$ and
its decay width is calculated in a model-independent way to be 0.75 --
0.21 MeV for $m_{\sg}=12$ -- 16 GeV \cite{Cheung:2002rk}. Additional
``sbottom splitting'' diagrams \cite{Malhotra:2003da} raise $\Gamma(Z \to
q {\bar q} \sg \sg)$ by about 0.01 MeV.

By studying the SUSY-QCD effects on $R_b$, Cao {\it et al.} 
\cite{Cao:2001rz} showed that the mass of the heavy bottom squark
should not exceed 195 GeV at the $3\sigma$ level. Taking into account
additional electroweak corrections to the gauge boson propagators and
including other relevant $Z$-pole observables and the $W$-boson mass, Cho
\cite{Cho:2002mt} found that the mass of $\sb'$ is further constrained
($\le 180$ GeV) and that one of the top squarks ($\st$) should be lighter
than 98 GeV at the $5\sigma$ level. However, both of these analyses
ignored the impact of the new $Z$ decays on the electroweak
observables. Most notably, it was shown in Ref.~\cite{Cheung:2002rk} that
$Z \to b\bar{\sb}\sg/{\bar b}\sb\sg$ may affect the measurement of $R_b$.
Indeed, since the gluino is assumed to decay promptly to a bottom quark
and a bottom squark, the final state in $Z \to b\bar{\sb}\sg/{\bar
b}\sb\sg$ will be an energetic bottom jet back-to-back with a ``fat'' jet
consisting of a bottom quark and two light bottom squarks. The energetic
bottom jet can add to $R_b$ if the analysis consists merely of counting
$b$ candidates. Moreover, if the light bottom squark cannot escape the
detector, the ``fat'' jet can also be tagged as a $b$ jet and contribute
to $R_b$ \cite{Cheung:2002rk,RH}.  However, since the branching ratio of
$Z \to b\bar{\sb}\sg/{\bar b}\sb\sg$ is only of order $10^{-3}$, current
LEP data are not likely to be sensitive to it \cite{RH} without a
dedicated search. Despite this, the $Z \to b\bar{\sb}\sg/{\bar b}\sb\sg$
decay plays a nonnegligible role in constraining the heavy bottom squark
mass. In this Letter we argue that the upper bound on $m_{\sb'}$ will be
relaxed if this process can contribute a certain amount to the measured
$\Gamma(Z \to b \bar b)$.

The proposed scenario implicitly requires that $R$-parity be violated in
bottom squark decays \cite{Cho:2002mt}. Assuming pure lepton-number
violation in top squark decays (${\cal B}(\st \to \tau b)=1$), the CDF
Collaboration \cite{Acosta:2003ys} has recently set a 95\% confidence 
level lower limit at 122 GeV on the light top squark mass $m_{\st}$ in the
framework of $R$-parity violating MSSM. We will show that the upper bound
on $m_{\st}$ can be raised above 122 GeV. However, we point out that the
CDF lower limit is somewhat irrelevant to the proposed scenario. The top
squark decay in the light gluino and light bottom squark scenario is not
necessarily $R$-parity violating or purely lepton-number-violating. An
$R$-parity-violating decay invoked for the bottom squark does not tell us
anything about how the top squark decays since the two processes arise
from different terms in the Lagrangian.

The $Z b {\bar b}$ vertex in the Standard Model can be written as
\be V^\mu(Z b {\bar b}) = - i
\frac{g}{2\cos\theta_W}\left[\gamma^\mu(g_V^b -
g_A^b\gamma^5)\right]~~, \ee
where $g_V^b = -1/2 + 2/3 \sin^2\theta_W$, $g_A^b = -1/2$, and $\theta_W$
is the weak angle. The SUSY-QCD corrections to $g_V^b$ and $g_A^b$ have
been calculated in the MSSM by a number of authors \cite{Cao:2001rz,
Baek:2002xf, Hagiwara:1990st, Djouadi:1993ix, Cho:1999km}. Using the
analytical expressions in Ref.~\cite{Djouadi:1993ix}, we reproduce the
numerical results in Ref.~\cite{Cao:2001rz}. However, we cannot reproduce
the numerical results in Ref.~\cite{Baek:2002xf}, which also disagree with
the other references analytically.

\begin{figure}
\begin{center}
\includegraphics[height=7in]{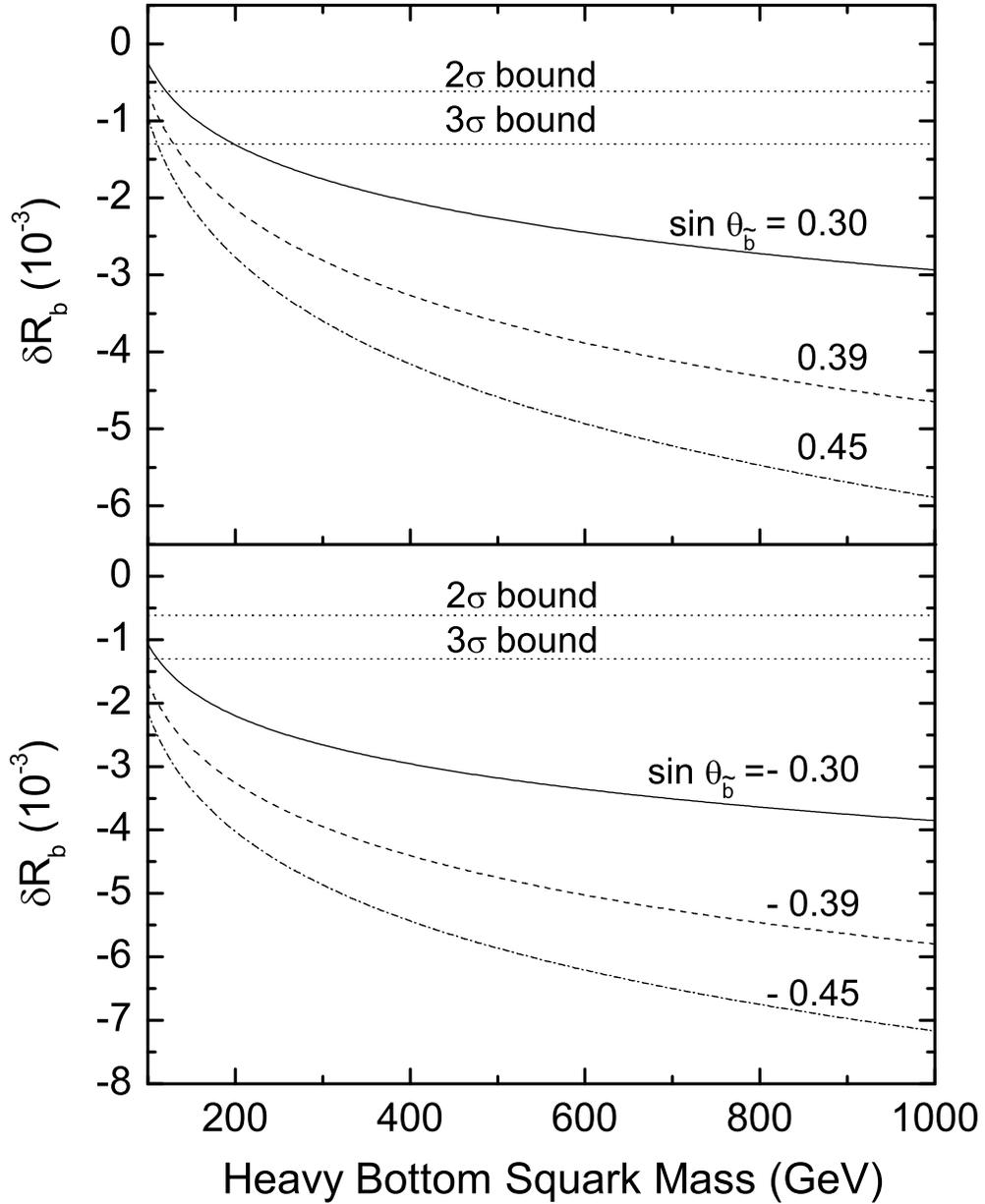} \\
\caption{Change in $R_b$ due to SUSY-QCD effects alone in the light gluino
and light bottom squark scenario with respect to its Standard Model value
as a function of the heavy bottom squark mass. Here we take $m_{\sb}=5.5$
GeV, $m_{\sg}=16$ GeV. The horizontal dotted lines correspond to $2\sigma$
and $3\sigma$ bounds set by the LEP $R_b$ measurements \cite{PDG}.
\label{fig:Rb}}
\end{center}
\end{figure}

For numerical evaluation, we take $\alpha_s(M_Z)= 0.120$, $\sin^2\theta_W
= 0.2311$, $m_b = 4.1$ GeV, $m_{\sb} = 5.5$ GeV, $m_{\sg}=16$ GeV,
$R_b^{\rm SM} = 0.21569$ and $\Gamma_Z^{\rm SM}({\rm had}) = 1.7429$
GeV \cite{PDG}. The greatest possible values are chosen for $m_{\sb}$ and
$m_{\sg}$ so that the upper limit on $m_{\sb'}$ is the least restrictive
\cite{Cao:2001rz}. The change in $R_b$ with respect to its SM value is
plotted in Fig.~\ref{fig:Rb}. For now we do not consider the possible
effects of the new $Z$ decay processes. The bottom squark mixing angle
$\theta_{\sb}$ is chosen to satisfy $|\sin\theta_{\sb}| \simeq 
\sqrt{2\sin^2\theta_W/3} \simeq 0.39$ so that the $Z\sb \bar {\sb}$
coupling is suppressed \cite{Berger:2000mp}. We allow $|\sin\theta_{\sb}|$
to vary between 0.30 and 0.45. The partial decay width $\Gamma(Z \to \sb
\bar {\sb})$ is about (1.0, 0.6) MeV for $|\sin\theta_{\sb}|$ =
(0.30,~0.45). Fig.~\ref{fig:Rb} shows that the heavy bottom squark mass
$m_{\sb'}$ must be less than about 200 GeV for $R_b$ to be within the
$3\sigma$ bound ($\delta R_b \ge -0.0013$) set by the LEP experimental
value \cite{PDG}.

\begin{figure}
\begin{center}
\includegraphics[height=4.5in]{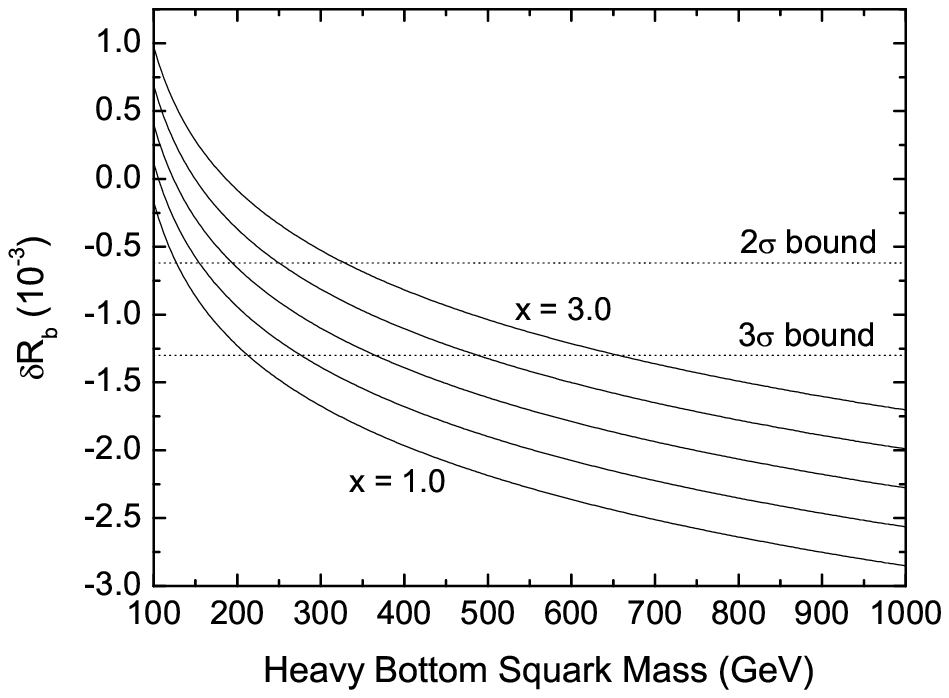} \\
\caption{Change in $R_b$ in the light gluino and light bottom squark
scenario with respect to its Standard Model value as a function of the
heavy bottom squark mass. SUSY-QCD effects and possible impact of the
decay process $Z \to b\bar{\sb}\sg/{\bar b}\sb\sg$ are combined. $x$ is
the contribution (in units of MeV) of the process to measured $\Gamma(Z
\to b \bar b)$ as a result of mistagging of the ``fat'' jets. From bottom
to top, x takes the values 1.0, 1.5, 2.0, 2.5 and 3.0. Here we use
$m_{\sb}=5.5$ GeV, $m_{\sg}=16$ GeV and $\sin\theta_{\sb} = 0.30$. The
horizontal dotted lines correspond to $2\sigma$ and $3\sigma$ bounds set
by the LEP $R_b$ measurements \cite{PDG}. \label{fig:UL}}
\end{center}
\end{figure}

Now suppose that a portion of the $Z \to b\bar{\sb}\sg/{\bar b}\sb\sg$
events cannot be distinguished from the $Z \to b \bar b$ ones and hence
contribute $x$ MeV to $\Gamma(Z \to b \bar b)$. We shall take $1.0 \le x
\le 3.0$. Note that the decay width for $Z \to b\bar{\sb}\sg/{\bar
b}\sb\sg$ is approximately 4.4 MeV for $m_{\sb} = 3$ GeV, $m_{\sg}=16$ GeV
and $\sin\theta_{\sb} \simeq 0.39$ \cite{Cheung:2002na}. Since the two
bottom quarks in the final states will hadronize in the detector, the
hadronic decay width $\Gamma_Z({\rm had})$ will also be affected.
Furthermore, the final-state bottom squark can also contribute to the
hadronic decay width via $R$-parity-violating decays to light quarks
\cite{Berger:2000mp}. (Possibilities of a long-lived $\sb$ and
lepton-number-violating decays of $\sb$ to hard leptons are both
disfavored; see Refs.~\cite{Berger:2000mp, Heister:2003hc}.) Considering
also other decays such as $Z \to \sb \sb$, $Z \to \sg \sg$ and $Z \to q
\bar q \sg \sg$, we estimate that the total change in $\Gamma_Z({\rm
had})$ is $4.0\ {\rm MeV} +\delta\Gamma(Z \to b\bar b)$. Here
$\delta\Gamma(Z \to b\bar b)$ denotes the change in $\Gamma(Z \to b \bar
b)$ due to SUSY-QCD effects, as discussed above. The upper bound on
$m_{\sb'}$ can then be relaxed if the additional decays contribute
positively to $R_b$, i.e., $[\Gamma^{\rm SM}(Z \to b \bar b) +
x]/[\Gamma^{\rm SM}_Z({\rm had})+4.0] > R_b^{{\rm SM}}$, or equivalently
$x/4.0>R_b^{{\rm SM}}$. For example, the new upper bound on $m_{\sb'}$ is
around 275 GeV at the $3\sigma$ level for $x = 1.5$; see 
Fig.~\ref{fig:UL}, where we take into account the additional $Z$ decays
and plot $\delta R_b$ as a function of $m_{\sb'}$ for $m_{\sb}=5.5$ GeV,
$m_{\sg}=16$ GeV, $\sin\theta_{\sb} = 0.30$ and $x =$ 1.0 -- 3.0. The new
upper bounds at the $3\sigma$ level are presented in Table~\ref{tab:UL}
for various values of $x$.

%

\begin{table}
\caption{Upper bounds on the heavy bottom squark mass at the $3\sigma$
level. $x$ is the contribution (in units of MeV) of the decays $Z \to
b\bar{\sb}\sg/{\bar b}\sb\sg$ to measured $\Gamma(Z \to b \bar b)$ as a
result of mistagging of the ``fat'' jets. \label{tab:UL}} 
\begin{center}
\begin{tabular}{c c} \hline \hline
$x$ & Upper bound on $m_{\sb'}$ (GeV) \\
\hline
1 & 215 \\
1.5 & 275 \\
2 & 365 \\
2.5 & 490 \\
3 & 655 \\
\hline \hline
\end{tabular}
\end{center}
\end{table}

SUSY-QCD corrections to $g_V^b$ and $g_A^b$ also affect some other related
$Z$-pole electroweak observables. These are $\Gamma_Z$, $\Gamma({\rm
had})$, $R_\ell$, $R_c$, $A_{\rm FB}^{(0,b)}$ and $A_b$, where we use the
notations in Ref.~\cite{PDG}. Some or all of these observables can also be
affected by the additional $Z$ decay processes in the proposed 
scenario. We show in Table~\ref{tab:obs} the comparison of the predictions
in the SM and in the proposed scenario for $m_{\sb'}=180,\ 275$ GeV, with
or without the additional decays taken into account. (Little can be known
of the effect of those decays on $A_{\rm FB}^{(0,b)}$ and $A_b$.) One can
see that raising $m_{\sb'}$ from 180 GeV to 275 GeV does not make the
predictions agree much worse with experiments, especially after the
additional decays are considered. Thus, the constraint on $R_b$ plays the
most important role in setting an upper bound on $m_{\sb'}$, which we have
already shown can be relaxed to about 275 GeV if the process $Z \to
b\bar{\sb}\sg/{\bar b}\sb\sg$ contributes 1.5 MeV (about one third of its
own decay width) to the measured $\Gamma(Z \to b \bar b)$. With $m_{\sb'}
\simeq 275$ GeV, we argue below that the upper bound on $m_{\st}$ can also
be relaxed.

\begin{table}
\caption{Deviations of the predictions for some $Z$-pole observables from
experimental values \cite{PDG}. MSSM$_{180,\ 275}$ (1) [(2)] denotes the
prediction in the light gluino and light bottom squark scenario for
$m_{\sb'}=180,\ 275$ GeV before [after] taking into account the additional
$Z$ decays. Here we use $m_{\sb}=5.5$ GeV, $m_{\sg}=16$ GeV,
$\sin\theta_{\sb}=0.30$.
\label{tab:obs}}
\begin{center}
\begin{tabular}{c c c c c c} \hline \hline
Observable & SM & MSSM$_{180}$ (1) & MSSM$_{180}$ (2) &
MSSM$_{275}$ (1) & MSSM$_{275}$ (2) \\
\hline $\Gamma_Z$ & $0.61\sigma$ & $-0.52\sigma$ & $1.87\sigma$ &
$-1.00\sigma$ & $1.39\sigma$ \\
$\Gamma({\rm had})$ & $-0.75\sigma$ & $-2.05\sigma$ &
$-0.05\sigma$ & $-2.60\sigma$ & $-0.60\sigma$\\
$R_\ell$ & $-1.20\sigma$ & $-1.82\sigma$ & $-0.87\sigma$ &
$-2.08\sigma$ & $-1.12\sigma$ \\
$R_c$ & $-0.19\sigma$ & $-0.11\sigma$ & $-0.23\sigma$ &
$-0.07\sigma$ & $-0.20\sigma$ \\
$A_{\rm FB}^{(0,b)}$ & $3.18\sigma$ & $3.39\sigma$ & -- &
$3.43\sigma$ &
-- \\
$A_b$ & $0.69\sigma$ & $0.85\sigma$ & -- & $0.88\sigma$ & -- \\
\hline \hline
\end{tabular}
\end{center}
\end{table}

The constraint on $m_{\st}$ originates primarily from $SU(2)_L$ gauge
symmetry and the constraint on oblique corrections, i.e., electroweak
corrections to gauge boson propagators. These corrections are parametrized
by $\Delta S_Z$, $\Delta T_Z$ and $\Delta M_W$ in 
Ref.~\cite{Cho:1999km}. It has been shown \cite{Cho:2002mt, Cho:1999km}
that the dominant contributions to these parameters come from the bottom
and top squark corrections (denoted by $\Delta T$) to the $T$-parameter
\cite{Peskin:1990zt}. Since $T$ is defined in terms of vacuum polarization
amplitudes of the $SU(2)_L$ gauge bosons, only the left-handed components
of the squarks can contribute to $\Delta T$. For given bottom squark 
masses $m_{\sb}$, $m_{\sb'}$ and left-right ($L$-$R$) mixing angle
$\theta_{\sb}$, large $L$-$R$ mixing for the top squarks is favored so
that the left-handed component of the lighter top squark mass eigenstate
is relatively suppressed and therefore a small $\Delta T$ can be obtained
\cite{Cho:2002mt}. The $L$-$R$ mixing for the top squarks is parametrized
by $A_{\rm eff}^t$ in Ref.~\cite{Cho:2002mt}, where it is constrained to
be no less than 300 GeV for $m_{\sb'} = 180$ GeV in the light gluino and
light bottom squark scenario. This leads to a very restrictive upper bound
(98 GeV) on $m_{\st}$. For $A_{\rm eff}^t = 300$ GeV and a consistent set
of masses and mixing angles ($m_{\sb} = 5.5$ GeV, $m_{\sb'} = 180$ GeV,
$m_{\st} = 98$ GeV, $m_{\st'} = 340$ GeV, $\theta_{\sb}=\arcsin 0.30
\simeq 17.5^\circ$, $\theta_{\st}=49.2^\circ$), we find $\Delta T \simeq
0.081$. Larger values for $\Delta T$ are obtained if $m_{\st} > 98$ GeV.
Without going through a similar $\chi^2$ fit procedure to that in
Ref.~\cite{Cho:2002mt}, we regard 0.08 as a threshold value for $\Delta T$
below which oblique corrections are assumed to be acceptably small. For
the same $A_{\rm eff}^t$ but with $m_{\sb'}$ raised to 275 GeV, we find
that smaller values for $\Delta T$ can be obtained for $m_{\st}$ as large
as 210 GeV; see Table~\ref{tab:T}. This implies that the upper bound on
$m_{\st}$ may be considerably relaxed if the process $Z \to  
b\bar{\sb}\sg/{\bar b}\sb\sg$ can contribute a certain amount to the
measured $\Gamma(Z \to b \bar b)$.

\begin{table}
\caption{Examples for $m_{\st} > 98$ GeV and $\Delta T < 0.08$. We
take $m_{\sb}=5.5$ GeV, $m_{\sb'}=275$ GeV,
$\sin\theta_{\sb}=0.30$. \label{tab:T}}
\begin{center}
\begin{tabular}{c c c c} \hline \hline
$m_{\st}$ (GeV) & $m_{\st'}$ (GeV) & $\theta_{\st}$ & $\Delta T$ \\
\hline
100 & 357 & $31.9^{\circ}$ & 0.061 \\
165 & 368 & $38.1^{\circ}$ & 0.070 \\
210 & 387 & $46.4^{\circ}$ & 0.079 \\
\hline \hline
\end{tabular}
\end{center}
\end{table}

One can even tolerate much larger values of $\Delta T$ if one is prepared
to consider heavier Higgs boson masses and small positive changes in $S$
in precision electroweak fits \cite{bigT,JRAPV}.  In one example
\cite{JRAPV}, by letting the Higgs boson mass rise to about 1 TeV, one can
accommodate $\Delta T$ as large as 0.5 at the price of $\Delta S = 0.3$
while retaining an acceptable fit to the data.

Finally we comment briefly on the scale-dependence (``running'') of the
strong coupling constant $\alpha_s$ in the light gluino and light bottom
squark scenario. In the context of this scenario, we showed in a previous
work \cite{Chiang:2002wi} that $\alpha_s(M_Z)$ fell in the range (0.130 --
0.135) $\pm$ $(>)$0.003 if extrapolated from low mass scales (such as
$m_b$) and in the range (0.123 -- 0.131) $\pm$ 0.005 if extracted directly
from the $Z$-pole observable $\Gamma_Z({\rm had})$. These two ranges
overlap with one another and no clear-cut decision could be made in favor
of either the Standard Model or the proposed SUSY scenario. However, we
were only able to evaluate $\Gamma(Z \to \sg\sg)$ very roughly and did not
take into account the decay process $Z \to b \bar{\sb}\sg/{\bar b}\sb\sg$
at that time. With both of these processes now well understood and
considered, the possible range of the extracted $\alpha_s(M_Z)$ from
$\Gamma_Z({\rm had})$ turns out to be different. For numerical purposes,
we again use $\alpha_s(M_Z)=0.120$, $m_{\sb}=5.5$ GeV, $m_{\sg}=16$ GeV,
$\sin\theta_{\sb}=0.30$ and take 120 GeV $\le m_{\sb'} \le 655$ GeV. Here
the lower bound on $m_{\sb'}$ is based on a recent analysis by Berger {\it
et al.} \cite{Berger:2003sn}. The SUSY-QCD correction to $\Gamma(Z \to b
\bar b)$ is then $\delta\Gamma(Z \to b \bar b)= - (1.3$ -- 5.6) MeV. We
neglect the relatively small changes in $\delta\Gamma(Z \to b \bar b)$ due
to different choices of $\alpha_s(M_Z)$. Assuming the contribution of the
new decay processes to $\Gamma_Z({\rm had})$ is 4 MeV, the net change in
the predicted $\Gamma_Z({\rm had})$ is then ($-1.6$ -- $+2.7$) MeV. This
change can be accounted for by tuning $\alpha_s(M_Z)$ to lie in the range
(0.118 -- 0.126) $\pm$ 0.005. It is lower than the extrapolated range from
low mass scales, but not at a statistically significant level. The heavy
sbottom mass should be better constrained to reduce the indeterminacy. 

To summarize, we have shown that it is possible to circumvent restrictive
upper bounds that have been placed on the heavy bottom squark mass when the
gluino and one bottom squark are both light.  The reduction of $R_b$
predicted in such treatments, which would conflict with data, can be
compensated by a contribution to the $b$ quark production cross section
from the process $Z \to b\bar{\sb}\sg/{\bar b}\sb\sg$.  In such a case
one expects a fast $b$ quark jet to be accompanied by a ``fat jet'' associated
with the hadronization of the $\bar{\sb}\sg$ or $\sb\sg$ system.  The
relaxation of the bounds on the heavy bottom squark mass is accompanied by a
corresponding relaxation of bounds on the top squark mass. 
The situation of the scale-dependence of the strong coupling constant
$\alpha_s$ becomes less favorable to the light gluino and light bottom
squark scenario after the decay process $Z \to b\bar{\sb}\sg/{\bar 
b}\sb\sg$ is taken into account. However, improved understanding of the
effects of the new decay processes in the scenario on $R_b$ and
$\Gamma_Z({\rm had})$ is needed for a clear-cut conclusion.

\section*{Acknowledgements}

We are grateful to S.~Baek, E.~L.~Berger, J.~Cao and R.~Hawkings
for helpful discussions. We would also like to thank D.~Rainwater
and T.~M.~Tait for useful comments. This work was supported in
part by the U.\ S.\ Department of Energy through Grant Nos.\
DE-FG02-90ER-40560.

\end{document}